# Detection of Generalized Synchronization using Echo State Networks

D Ibáñez-Soria[1], J Garcia-Ojalvo[2], A Soria-Frisch[1], G Ruffini[1,3]
[1]Starlab Barcelona S.L., Neuroscience Research Business Unit
[2] Department of Experimental and Health Sciences, Universitat Pompeu Fabra, Barcelona, Spain
[3]Neuroelectrics Corporation
E-mail: david.ibanez@starlab.es

**Abstract.** Generalized synchronization between coupled dynamical systems is a phenomenon of relevance in applications that range from secure communications to physiological modelling. Here we test the capabilities of reservoir computing and, in particular, echo state networks for the detection of generalized synchronization. A nonlinear dynamical system consisting of two coupled Rössler chaotic attractors is used to generate temporal series consisting of time-locked generalized synchronized sequences interleaved by unsynchronized ones. Correctly tuned, echo state networks are able to efficiently discriminate between unsynchronized and synchronized sequences. Compared to other state-of-the-art techniques of synchronization detection, the online capabilities of the proposed ESN based methodology make it a promising choice for real-time applications aiming to monitor dynamical synchronization changes in continuous signals.

## 1. Introduction

In its everyday use, the concept of synchronization is commonly taken to imply identical behavior between two interacting systems. However, when the affected systems are chaotic, more sophisticated synchronization forms can exist, such as phase synchronization (where only the phase but not the amplitude of two chaotic oscillations agree with each other) [1], lag synchronization (where one of the chaotic systems follows the other with a certain delay) [2], or generalized synchronization (where the states of the two systems are functionally –but not identically– related in a nontrivial, in general nonlinear, manner) [3,4]. Generalized synchronization, in particular, has proven to be a relevant feature in the analysis of neurological disorders such as Alzheimer's disease [5] and brain tumors [6].

Over the years, several methods have been proposed for the detection of generalized synchronization, including the replica method [7], the synchronization likelihood approach [8], and the mutual false nearest neighbor method [4], among others. These methods are usually computationally costly, cannot be applied in a continuous manner, and in some cases, suffer from different biases [9]. Here we propose a machine-learning-based approach that enables the online detection of generalized synchronization in an effective manner. The method relies on a recurrent neural network to provide the necessary fading memory that allows processing dynamical signals.

Unlike feedforward neural networks, in which a static input-output mapping is applied, recurrent neural networks (RNNs) have cyclic connections that provide memory, implementing a system with dynamical capabilities [10]. Training RNNs has traditionally been computationally more expensive than training feedforward networks. Echo State Networks [11] (ESN) and Liquid State Machines [12] (LSM), which were developed independently and simultaneously, and the more recent decorrelation learning rule for RNNs [13], are complementary approaches for designing, training and analyzing RNNs within a methodological framework known as Reservoir Computing (RC). RC is based on the principle that if the network possesses certain algebraic properties the supervised training of all connections is not necessary. Only supervised training of readout weight is sufficient to obtain optimal classification performance in many tasks.

Here we examine the capabilities of ESNs in the task of detecting generalized synchronization changes in synthetic temporal series based on the coupling of two chaotic systems. We will use in particular the well-known Rössler attractors. The paper is structured as follows. In section 2 we explain the methodology followed to construct generalized synchronized sequences using coupled chaotic Rössler oscillators. In section 3 we provide an overview of the echo state network architecture and their key training parameters. The procedure used to study ESN generalized synchronization detection is presented in section 4 and the obtained results in section 5. We conclude with a discussion in section 6.

## 2. Generation of in silico time series

The main objective of this work is to explore the capabilities of ESNs for discriminating between generalized synchronized time-series and unsynchronized sequences. In this section, we describe how synchronized chaotic attractors are constructed. To that end we follow the unidirectional (master-slave) coupling of two Rössler oscillators proposed by Rulkov et al [4]. A Rössler oscillator is a dynamical system defined by three non-linear ordinary differential equations that exhibit chaotic dynamics. The Rössler oscillator described by the state variables $x$ presented in (1) has been adopted as the driving system, while $y$ in (2) constitutes the coupled response system:

$$\dot{x}_1 = -(x_2 + x_3)$$
$$\dot{x}_2 = x_1 + a\, x_2$$
$$\dot{x}_3 = a + x_3(x_1 - b)$$
(1)

$$\dot{y}_1 = -(y_2 + y_3) - g(y_1 - x_1)$$
$$\dot{y}_2 = y_1 + 0.2\, y_2$$
$$\dot{y}_3 = 0.2 + y_3(y_1 - 5.7)$$
(2)

The manifold $x_1 = y_1$, $x_2 = y_2$ and $x_3 = y_3$ contains the trajectories of synchronized oscillations. The conditional Lyapunov exponents (CLE) of the response system characterize its asymptotic local stability [14]. It has been proven that if all CLEs are negative, the response system is stable and shows synchronization [15]. Reservoir computing has proved its capabilities for the assessment of positive and negative Lyapunov exponents in high dimensional spatio-temporal chaotic systems [16].

For a = 0.2 and b = 5.7 Rulkov et al [4] proved that trajectories of synchronized motions were stable for a coupling factor of g = 0. 2, becoming unstable for g = 0.15.

Figure1A shows a plot of $x_2(t)$ vs $y_2(t)$ for $g = 0.2$, and Figure 1B for $g = 0.15$. In A we observe a straight line that denotes identical synchronization between driving and response system, while in B unsynchronized oscillations appear. By correctly tuning the coupling factor g it is thus possible to force synchronization between the coupled systems. To achieve generalized synchronization Rulkov proposes the nonlinear transformation of the response system $y(t)$ into $z(t)$ presented in equation 3.

$$z_1 = y_1$$
$$z_2 = y_2 + 0.4y_3 + 0.03y_3^2 \quad (3)$$
$$z_3 = y_3$$

Figure1C plots $x_2(t)$ vs $z_2(t)$ for $g = 0.2$, where a straight line denoting identical synchronization is no longer observed. Figure1D shows $x_2(t)$ vs $z_2(t)$ for $g = 0.15$. In C, even though a complex relationship between variables is observed, synchronization is not lost since only a non-linear transformation was applied. In this case $x_2(t)$ and $z_2(t)$ present generalized synchronization [4].

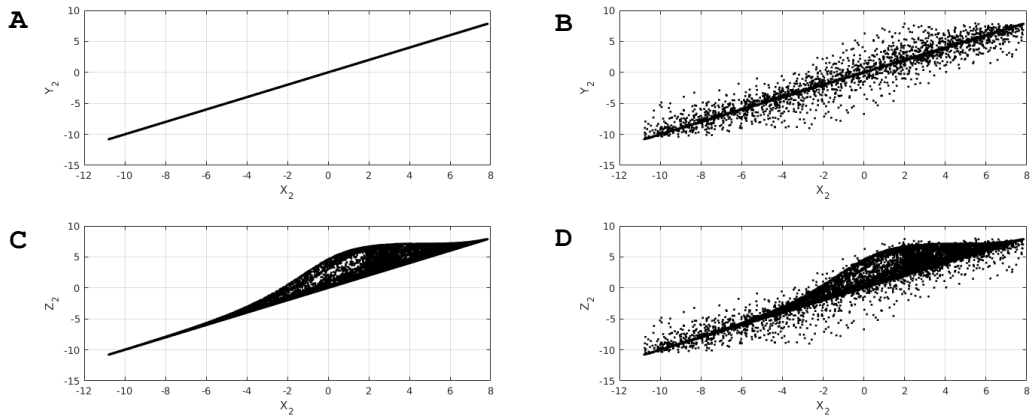

**Figure1** Dynamics of two coupled Rössler oscillators: (**A**) $x_2(t)$ vs $y_2(t)$ for $g = 0.2$, (**B**): $x_2(t)$ vs $y_2(t)$ for $g = 0.15$, (**C**) $x_2(t)$ vs $z_2(t)$ for $g = 0.2$ and (**D**) $x_2(t)$ vs $z_2(t)$ for $g = 0.15$.

Using the behavior reported in Fig. 1, a continuous temporal signal consisting of a series of synchronized sequences interleaved by unsynchronized ones has been constructed. For this purpose, a the time-varying square wave coupling function g(t) is used:

$$g(t) = 0.05 \left(\frac{1}{2} sign\left(\sin\left(\frac{pi*t}{T}\right)\right) + 1\right) + 0.15 \quad (4)$$

where T defines the up or down state duration in time units. The model has been solved using a 2$^{nd}$-3$^{rd}$ order Runge-Kutta method (implemented with the Matlab function ode23) with fixed integration step of 0.4 time units. The synchronization detection performance is evaluated using time series of length $10^6$ time units and an up/down state duration T of $10^5$ time units. The resultant signal thus concatenates, depending on the initial time, at least 5 generalized synchronized sequences followed by 5 unsynchronized ones.

## 3. ESNs for generalized synchronization detection

We now discuss the ESN architectures explored, and the tests to assess generalized synchronization. Artificial Neural Networks (ANN) such as the multilayer-perceptron (MLP) present a feed-forward structure where the information flows from the input nodes, through the hidden nodes and to the output nodes [17]. Based on the Representation Theorem, ANNs (and in particular MLPs) are able to approximate any given arbitrary function. This static input-output architecture makes them suitable for the analysis of stationary problems, but it is in general not adequate to deal with dynamical time-dependent problems. Recurrent neural networks (RNNs) incorporate cyclic connections that allow the system to provide memory capability to the network, and therefore to encode time-dependent information. This addition transforms the network into a dynamical system. The network keeps in its internal states non-linear transformations of the input history (fading memory) allowing it to process information with temporal context [18].

RNNs training have traditionally been computationally expensive to train because of their cyclic non-linear nature [22]. Reservoir Computing (RC), a methodological framework to understand, train, and apply Recurrent Neural Networks (RNNs), was proposed independently and simultaneously with the development of Echo State Networks (ESNs) [11] and Liquid State Machines (LSMs) [19]. The fundamental principle of ESN is that if the network presents a certain algebraic property known as the Echo State Property (ESP), only supervised training of readout connections is needed [20]. Trained output units combine the internal RNN dynamics into desired outputs. The untrained RNN is called the dynamical reservoir (DR) and is formed by input, hidden and backpropagation connections. The echo state property ensures that the reservoir state does not depend in the long term on the initial conditions, which are thus forgotten as time passes [21].

The precise values of input and interior weights are irrelevant to the echo state property. These weight values can be randomly generated according to some parameters [22], among which it is worth pointing out the so-called spectral radius. The spectral radius, calculated as the largest absolute eigenvalue of the internal connections matrix, determines the timescale of the reservoir. In practice, if the spectral radius is smaller than one, the echo state property holds for most applications [22]. Some unlikely exceptions to this rule have been proposed, however [23]. Additionally, in some situations the echo state property may also hold for values of the spectral radius larger than one [21].

The spectral radius and some other key parameters of the ESN, mainly the size of the reservoir and the input scaling, rule the dynamical behavior of the network [22]. A small spectral radius induces a faster response, while a larger spectral radius is more suitable for tasks requiring longer fading memory. In general, the size of echo state networks can be larger compared to other recurrent neural network approaches [24]. In general and given sufficient data, the larger the network size, the better it can learn complex dynamics. In case of data shortage, large reservoirs can lead to overfitting and make the network present poor predictive performance. In ESNs, input scaling is implemented multiplying every input sample by the same scaling factor. The input scaling determines the degree of nonlinearity in the reservoir: while linear tasks require small input scaling factors, tasks with complex dynamics demand larger input scaling values.

In our case, we configured a single ESN network to have two input nodes and a single output node. The signals $x_2(t)$ and $z_2(t)$ from the Rössler oscillator feed the input nodes. The optimal parameters of the ESN have been determined through exhaustive search in a grid. Concretely we have used following value grids: number of internal units - (5, 10, 25, 50, 75, 100, 200, 300, 400, 500), spectral radius - (1, 0.7, 0.5, 0.2, 0.1, 0.01, 0.001), and input scaling - (0.001, 0.01, 0.1, 0.5, 1, 5, 10, 25, 50, 75, 100). The number of internal units has to be large enough for the system to learn the complex dynamics associated with generalized synchronization, but not too large so the system generalizes well enough. The target of the ESN is to discriminate between synchronized and unsynchronized sequences. For this binary classification problem, during training the ESN output of synchronized sequences is teacher-forced to 1 and to -1 for unsynchronized sequences.

## 4. Generalized Synchronization Detection

The performance of each input scaling, network size and spectral radius tuple has been individually assessed. A Rössler signal formed by a series of 5 generalized synchronized sequences followed by unsynchronized sequences, as described in Section 2, is used for training. A different signal of the same characteristics and the same number of samples but starting at different initial time is used for testing the tuple performance. Hence we are using for performance evaluation a hold-out validation scheme with 50% of training samples and 50% of different test ones. According to the chaotic nature of the Rössler dynamics, by starting at different instants, although the two attractors present a common pattern in space state, they will develop trajectories that exponentially separates in an unsynchronized manner [25]. The ESN test output has been smoothed using a 10000-sample moving average window before performance evaluation.

The Receiver Operating Characteristic (ROC) Curve has been largely used as an effective method to evaluate the performance of binary classification systems. In such two-class prediction problems outcomes are labeled as positive and negative class. ROC curves show the trade-off between sensitivity and specificity as function of a varying decision threshold. In our case, the output test samples corresponding to synchronized sequences are labeled as the positive class and unsynchronized ones as the negative class. The false positive rate (specificity) and true positive rate (sensitivity) of the ESN output is then calculated for the all possible values of the decision threshold as applied on the ESN output after smoothing. Figure 3C exemplary shows the ROC curves calculated for different smoothing window lengths. The area under the ROC curve, or simply Area Under the Curve (AUC), measures the probability of the system of right ranking the positive and negative class samples. An area of 1 means that all samples were correctly classified while an area of 0.5 represents random classification. With the objective of reducing the random effect introduced by the reservoir initialization, each tuple performance evaluation process is repeated 5 independent times and its average AUC is used to assess the tuple's discrimination performance.

The largest AUC value (0.85) was found for a spectral radius equal to 0.01, an input scaling of 25, and 500 internal units. The reservoir size appears to be a key training parameter. Figure 2A displays the AUC score of the best spectral-radius/input-scaling tuple as function of the number of internal units. The AUC score increases asymptotically with the reservoir size. According to these results, it is necessary to have at least 100 internal units to achieve an AUC performance larger than 0.8. The performance of more than 500 internal units was not evaluated due to computational restrictions. The input scaling constitutes also a training parameter that sensitively affects performance. Figure 2B represents the AUC score for the best spectral radius as a function of the input scaling. The AUC performance significantly improves in the range [5, 50]. This large input scaling factor suggests a large non-linearity of the regression problem under evaluation. On the other hand, in this generalized synchronization detection scenario the discrimination performance robustly behaves with respect to the spectral radius as shown in Figure 2C, where the AUC score obtained for the best input scaling is plotted as a function of the spectral radius. Figures 2D, 2E and 2F corroborate the aforementioned observations, with the colour map representing the AUC for 100, 300 and 500 internal units respectively.

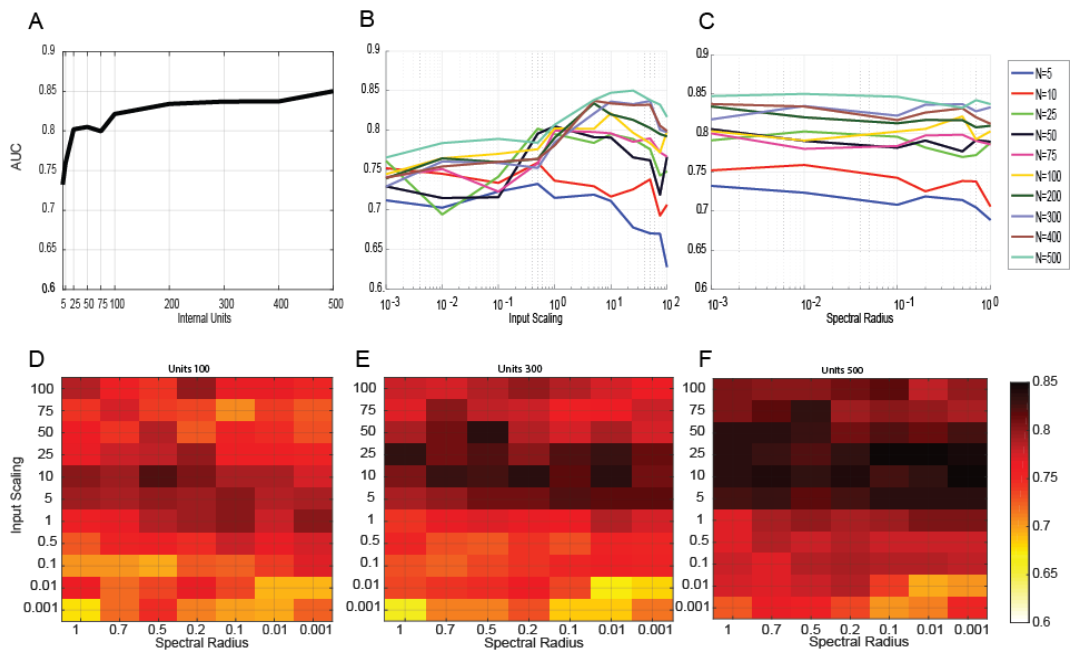

**Figure 2** Generalized synchronization detection performance parameter analysis. (A) AUC score of the best spectral-radius/input-scaling tuple as function of the reservoir size. (B) AUC score as function of the input scaling (C) AUC score as a function of the spectral radius. AUC (colormap value) variation with respect to spectral radius (X-axis) and input scaling (Y-axis) for different reservoir dimensions: (D) 100 units, (E) 300 units, and (F) 500 units.

Figure 3A depicts an example of the output of the outperforming ESN parameter tuple before averaging, which achieves a 0.54 AUC. The output has been scaled in the [-1, 1] for the sake of visualization. The dotted black line represents the aimed ESN Output, where 1 represents synchronized sequences and -1 unsynchronized ones. Despite of the small AUC obtained, a substantial difference between synchronized and unsynchronized intervals can be observed. Desynchronized samples present a larger high-frequency amplitude response. In Figure 1B, where the coupled attractors are considered to be unsynchronized, we can observe that many samples lie around the straight line corresponding to synchronization. The ESN discrimination of these sequences appears to be more difficult, as within them not all samples seem to be totally unsynchronized. This behaviour motivates the use of techniques that smooth the ESN output, in order to improve the detection performance. To smooth the ESN output we have used a simple moving average approach implemented as the unweighted mean of the previous W samples, where W states for the window length in number of samples. Figure 3B shows the ESN output after W=10000 samples averaging. In this case we can observe a better discrimination between generalized synchronized and unsynchronized sequences achieving an AUC of 0.85. Figure 3C shows the receivers operator curve computed for W = 1, 100, 500, 1000, 5000 and 10000 samples. As expected, the AUC increases with the averaging window length.

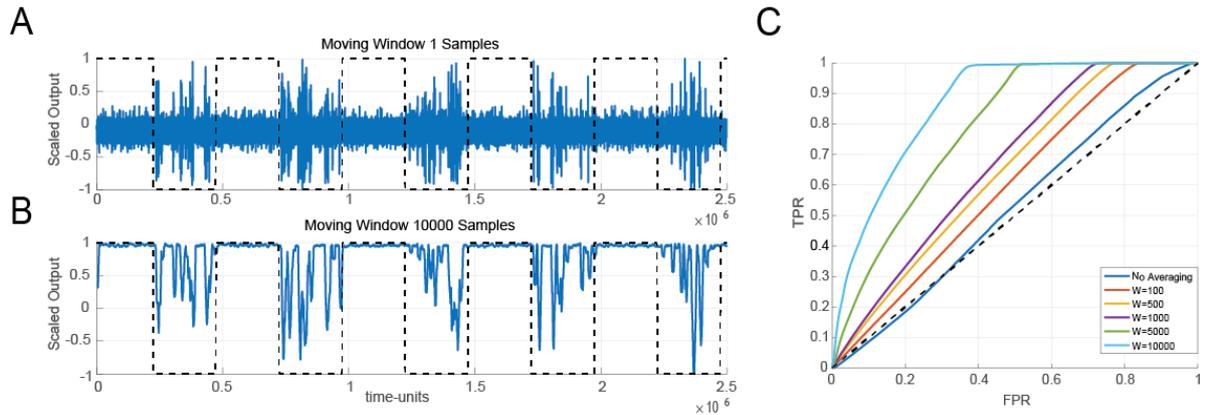

**Figure 3** (A) Unfiltered ESN output for spectral radius 0.01, input scaling 25, and 500 internal units (B) ESN output after 1000 samples moving average for spectral radius 0.01, input scaling 25 and 500 internal units. (C) Receivers operators curve calculated for averaging windows of 1, 500, 1000, 5000 and 10000 samples.

## 5. Discussion and Conclusions

We have presented a reliable, light training generalized synchronization detection methodology based on echo state networks. An optimal parameterization of ESNs was able to discriminate between time-locked generalized synchronized sequences from unsynchronized ones delivering an area under the curve above 0.85. Unlike other GS detection methods that cannot be applied in a continuous fashion, artificial neural networks update its output with every input sample. ESN therefore prove to be an ideal choice to develop applications capable of monitoring generalized synchronization changes in real-time.

An appropriate tuning of ESN parameters has proved necessary for achieving a good discrimination performance. The reservoir size turns out to be a fundamental training parameter along with the input-scaling. According to our results, a minimum of 100 internal units is required to achieve good performance and thus learning the generalized synchronization complex dynamics between Rössler oscillators. Input scaling determines the degree of non-linearity in the reservoir. An input scaling between 5 and 50 improved ESN discrimination capabilities. This large input scaling factor suggests the expected high non-linearity nature of generalized synchronization between the two coupled chaotic systems. We expect from the theoretical study presented herein to better understand the properties of ESN and the role of its different parameters. This can increase the number of applications of this ANN approach for the analysis of time series. In this context we are aiming to apply ESN for the analysis of electroencephalography data, which is used for brain monitoring, as we will show in future communications.

## 6  Acknowledgments

We want to thank the European Union's Horizon 2020 research and innovation programme who funded STIPED project under grant agreement No 731827. We also want to thank the support from ICREA Academia program and the Spanish Ministry of Economy and Competitiveness and FEDER (project FIS2015-66503-C3-1-P and Maria de Maeztu Programme for Units of Excellence in R&D, MDM-2014-0370). We also kindly thank developers of Oger toolbox.